\documentclass [12pt, a4paper]{article}
\usepackage{amssymb}
\usepackage{graphicx}
\usepackage{amsmath}
\begin{document}

\vspace*{40mm}

\centerline{\bfseries PROTECTED BARYON NUMBER  IN MODELS}  \par
\centerline{\bfseries WITH MAJORANA NEUTRINO } \par
\vspace{7mm}
\centerline{K. HASEGAWA }
\par
\centerline{\itshape Department of Physics, University of Alberta, Edmonton,} 
\centerline{\itshape Alberta, T6G 2J1, Canada}
\vspace{15mm}
\begin{abstract}
{We obtain the condition to protect the baryon number in the 
$SU(2)_{L}$ triplet Higgs model. The obtained condition gives
the upper limits for either of the two kinds of the coupling constants 
which are newly introduced in addition to ones in the standard model.}
\end{abstract}

\vspace{5mm}

\section{Introduction}
Neutrino oscillation experiments reveal that neutrinos have very small but finite masses. Since neutrinos have no
masses in the standard model,  we need to extend the standard model to induce the neutrino masses 
which coincide with the results of the neutrino oscillation experiments. It is possible that neutrino can
have both of Dirac mass and Majorana mass. Majorana mass has the feature that it violates the lepton number explicitly, which is conserved at classical level in the  standard model. Here we consider the case that neutrinos have Majorana masses. When we minimally extend the standard model within the framework of the $SU(2)_{L} \times U(1)_{Y}$ gauge theory for neutrino mass generation, typical three models exist, which are seesaw\cite{Yan,Yan2,Yan3}, Zee\cite{aZee}, and $SU(2)_{L}$ triplet Higgs model\cite{Sch}.

\par

The baryon number in the present universe is well known to be
\cite{Wmap1}
\begin{eqnarray}
B_{0}\equiv \frac{n_{B}}{s}=(8.4 \sim 9.1) \  \times 10^{-11}, 
 \end{eqnarray}
where $n_{B}$ is the baryon number density and $s$ is the entropy density. 
Since it is natural to think that there was no asymmetry between particle and anti-particle
at the very beginning of the universe, it is desired to search for the models or scenarios 
which generate the baryon number newly under the initial condition of the vanishing baryon number.
Fukugita-Yanagida scenario is one of the most reliable candidate for the baryon number generation \cite{Fuk}.
Fukugita-Yanagida scenario is based on the seesaw model which has the heavy right-handed Majorana neutrinos.
The out-of-equilibrium decay of the right-handed Majorana neutrinos produces the lepton number.
There exists spharelon-anomaly process in the standard model, which violates both of the baryon
and lepton numbers but conserves the difference between the baryon and lepton numbers.
The spharelon-anomaly process is in equilibrium at the temperature region between $100$ GeV and 
$10^{12}$ GeV. This process converts the produced lepton number into the baryon number
during the in-equilibrium region. It is well known that the Sakharov's three conditions need to be 
satisfied in order to generate the baryon number newly under the initial condition of the vanishing 
baryon number. 
The Sakharov's three conditions are the existence of a baryon number violation process,
CP violation, and an out-of-equilibrium process. In Fukugita-Yanagida scenario, in order to 
generate the lepton number newly, there exist the out-of-equilibrium decay of the 
right-handed Majorana neutrino which violates lepton number and CP phases
which  are accommodates with the decay processes. In such way Fukugita-Yanagida scenario
satisfies the Sakharov's three conditions. 

\par

In general it is possible that the models which can generate the baryon number newly under the
initial condition of the vanishing baryon number wash out the initial or the existing
baryon number. This feature in baryogenesis scenario is sometimes called `double-edged sold'.
In my research I focus on the washing out side in the double-edged sold. I assume that
the initial baryon number existed enough, for example, to the amount of the order unity, 
$B_{initial} \sim 1$. Then we focus on the models or parameter region in  a model which can
wash out the initial baryon number but can not generate one newly.
The purpose of my research is to find the condition to protect the initial baryon number
in Majorana neutrino model. I concretely find the condition to protect the baryon number
in the three models which I mentioned above,  seesaw\cite{Has1}, Zee\cite{Has2}, and $SU(2)_{L}$
triplet Higgs model\cite{Has3}. Since among these models the seesaw model can generate 
the baryon number newly by Fukugita-Yanagida scenario and the Zee model may generate
one by Electroweak baryogenesis scenario. In order to make these two models 
the ones which can wash out the initial baryon number but can not generate the
baryon number newly, I assume no CP violation at lepton sector in these models
and obtain the  condition to protect the initial baryon number.
On the other hand, the minimal version of $SU(2)_{L}$ triplet Higgs model which I use
can wash out the initial baryon number but can not generate the baryon number newly
being independent of the magnitude of CP violation\cite{Ma}. Thus this model is most suitable for
my standing point where I want to focus on the washing-out feature of Majorana neutrino models.
In the present proceeding we focus on the analysis in $SU(2)_{L}$ triplet Higgs model and obtain the
condition to protect the initial baryon number in the model.
In the second section we introduce the $SU(2)_{L}$ triplet Higgs model,
in the third section we obtain the condition to protect the initial baryon number,
and in the last section we have a summary.

\section{$SU(2)_{L}$ Triplet Higgs Model   \label{trimodel}}
In order to induce Majorana mass term of neutrinos in the $SU(2)_{L}$ triplet Higgs  model,
we add to the standard model two kinds of new  interactions which violate
the lepton number explicitly.
We newly introduce the $SU(2)_{L}$ triplet Higgs fields $\Delta$ assigned with 
the hypercharge $Y=2$  in addition to the fields of the standard model as
 \begin{eqnarray}
\Delta \equiv \left(\begin{array}{cc}
                                    \xi^{+}/\sqrt{2}  &   \xi^{++}    \\
                                     \xi^{0}  & -\xi^{+}/\sqrt{2}  \label{ddef}
									 \end{array}\right).
\end{eqnarray}
The Yukawa interactions of the triplet Higgs with the leptons are written as 
\begin{eqnarray}
 {\cal L}^{yukawa}_{\nu}
 &=& -\frac{1}{2}f^{\alpha \beta}Tr[T_{l^{\alpha},l^{\beta}}\Delta]
 +\mbox{h.c.}  \label{lld} \\ 
 &=& -\frac{1}{2}f^{\alpha \beta}\biggl[ \overline{(\nu^{\alpha})^{c}}\nu^{\beta}\xi^{0}-\frac{1}{\sqrt{2}}
   (\overline{(\nu^{\alpha})^{c}}e^{\beta}
   +\overline{(e^{\alpha})^{c}}\nu^{\beta})\xi^{+}-\overline{(e^{\alpha})^{c}}e^{\beta}\xi^{++}\biggr]
+\mbox{h.c.},  \nonumber
\end{eqnarray}
where $T_{l^{\alpha},l^{\beta}}$ is a $SU(2)_{L}$ triplet which is composed
of the two lepton doublets,
 \begin{eqnarray}
T_{l^{\alpha},l^{\beta}} \equiv \left(\begin{array}{cc}
                                    -\overline{(\nu_{L}^{\alpha})^{c}}e_{L}^{\beta}  & 
									\overline{(\nu_{L}^{\alpha})^{c}}\nu_{L}^{\beta}     \\
                                     -\overline{(e_{L}^{\alpha})^{c}}e_{L}^{\beta}  & 
									 \overline{(\nu_{L}^{\alpha})^{c}} e_{L}^{\beta}
									 \end{array}\right),
\end{eqnarray}
and $f^{\alpha \beta}$ is the symmetric Yukawa coupling constants ($f^{\alpha \beta}=f^{\beta \alpha}$). Since we take
the base where the mass matrix of the charged leptons is diagonalized,
the indices, $\alpha$ and $\beta$ take $e, \mu,$ and $\tau$.
However, when we assign the triplet Higgs fields 
with the lepton number $L=+2$, all the interactions in the model do not violate 
the lepton number, that is, the lepton number is not explicitly violated.
In order to violate the lepton number explicitly, we further introduce the cubic interaction of
the triplet Higgs with ordinary Higgs doublet 
$\Phi=(\phi^{+}, \phi^{0})^{t}$,
\begin{eqnarray}
 {\cal L}^{cubic}&=&-\frac{1}{2}A Tr[T_{\Phi,\Phi}\Delta^{\dagger}] +\mbox{h.c.}   \label{ppd5} \\
   &=& -\frac{1}{2}A \biggl[ (\phi^{+})^{2}\xi^{--}-\sqrt{2}
 \phi^{+}\phi^{0}\xi^{-}-(\phi^{0})^{2}\xi^{0 \ast}\biggr]+\mbox{h.c.}, \nonumber
\end{eqnarray}
where $T_{\Phi,\Phi}$ is a $SU(2)_{L}$ triplet which is composed
of the product of the two Higgs doublets,
\begin{eqnarray}
T_{\Phi,\Phi} \equiv \left(\begin{array}{cc}
                                    -\phi^{+}\phi^{0}  & \phi^{+}\phi^{+}   \\
                                     -\phi^{0}\phi^{0}  & \phi^{+}\phi^{0}
									 \end{array}\right), \label{phiphi}
\end{eqnarray}
which has the hypercharge $Y=+2$ and the lepton number $L=0$, and $A$ is the coupling constant.
We here assume the following Higgs potential,
\begin{eqnarray}
V(\Phi, \Delta)=-\mu^{2}\Phi^{\dagger}\Phi+\lambda(\Phi^{\dagger}\Phi)^{2} 
                            +M^{2}Tr[\Delta^{\dagger}\Delta]
							+\frac{1}{2}\bigl(A \ Tr[T_{\Phi,\Phi}\Delta^{\dagger}] +\mbox{h.c.}\bigr),    \label{pot} 
\end{eqnarray}
where $\mu^{2}$ and $M^{2}$ are positive.

\section{The Condition to Protect Baryon Number }
Whether the baryon number is protected or washed out depends on whether the processes which
violate baryon and lepton numbers are out of equilibrium or in equilibrium
under the equilibrium spharelon process.
We can estimate the out of equilibrium condition for a process as follow.
That a process is out of equilibrium means that the process can be regarded as vanishing in the finite age of universe, that is, the time scale of a process is longer than the age of the universe. That is written in 
the inequality,
\begin{eqnarray}
\frac{1}{H} \ < \ \frac{1}{\Gamma} \ \ \Leftrightarrow \ \  \Gamma \ < \ H,
\end{eqnarray}
where $\Gamma$ is the decay rate or interaction rate of a process and $H$ is Hubble parameter,
which is in radiation dominant epoch
\begin{eqnarray}
H=1.66 \sqrt{g_{\ast}} \frac{T^{2}}{M_{Pl}}=\frac{1}{2t}.  
\end{eqnarray}
with $g_{\ast}$ is the total degrees of freedom of effectively massless particles, $M_{Pl}$ being 
the Plank mass, and $t$ is the age of the universe. 
When some processes which violate the baryon and lepton numbers come in equilibrium, 
the baryon and lepton numbers are both washed out because the state where the number of particle 
is equal to the number of anti-particle has the maximal entropy.
On the other hand when some processes which violate the baryon or lepton number has been 
always out of equilibrium and there is any approximate conserved charge which includes 
the baryon number, the baryon number is protected and left in proportion to the initial value of
the approximately conserved charge. We next concretely find the condition to protect the baryon 
number in the $SU(2)_{L}$ triplet Higgs model.
At first we neglect the difference of the lepton flavor, $e, \mu, \tau$ for simplicity.
If the process which are caused through the interaction (\ref{lld}) are out of equilibrium as
\begin{eqnarray}
\Gamma(\xi^{0} \leftrightarrow \bar{\nu}_{L} +\bar{\nu}_{L}) \ < \ H,  \label{con1}
\end{eqnarray}
the $U(1)$ charge,
\begin{eqnarray}
P=B-L+2N_{\Delta}
\end{eqnarray}
is approximately conserved. Here $N_{\Delta}$ is triplet Higgs number.
If the process which are caused through the interaction (\ref{ppd5}) are out of equilibrium as
\begin{eqnarray}
\Gamma(\xi^{0} \leftrightarrow \phi^{0} + \phi^{0}) \ < \ H,  \label{con2}
\end{eqnarray}
the $U(1)$ charge,
\begin{eqnarray}
P=B-L
\end{eqnarray}
is approximately conserved. The conditions (\ref{con1}) and (\ref{con2}) become most sever when 
the temperature of the universe is equal to the scale of the triplet Higgs mass, $T \simeq M$.
Thus we obtain the condition to protect the baryon number as follow
\begin{eqnarray}
\Gamma(\xi^{0} \leftrightarrow \bar{\nu}_{L} +\bar{\nu}_{L}) \ < \ H \ |_{T=M}
\ \ 
\mbox{or}
\ \ 
\Gamma(\xi^{0} \leftrightarrow \phi^{0} + \phi^{0}) \ < \ H \ |_{T=M}.  \label{res}
\end{eqnarray}
The above derivation of the condition to protect the baryon number is 
a qualitative one. We can get a quantitative condition by solving the Boltzmann equation
in $SU(2)_{L}$ triplet Higgs model. I solve the Boltzmann equation and obtain
 the condition to protect the baryon number
which coincides with the condition (\ref{res}).
When we fix the mass of the triplet Higgs fields, the condition  (\ref{res}) gives
the upper limits for either of the two coupling constants, $f^{\alpha \beta}$ and $A$.
When we consider the difference of the lepton flavor, $e, \mu, \tau$,
 the condition to protect the baryon number (\ref{res}) is modified into the condition,
\begin{eqnarray}
\mbox{`At least one of the ten conditions in Table \ref{ap} must be satisfied'.} \label{res2}
\end{eqnarray} 
Here we omit the detailed discussion about this condition which is written in \cite{Has3}.

\begin{table}[ht] \begin{tabular}{|c|l|l|}\hline
 &  \ Approximately Conserved Charge & \ Condition \\ \hline 
(0)  & \ \ $P_{0}=B-L+2N_{\Delta}$  & \   $K_{A}<1$ \\ \hline 
(1)  & \ \  $P_{1}^{a}=\frac{B}{3}-L_{e}$  & \  $K_{L_{e}}<1$  \\ \hline 	
(2)  & \ \  $P_{1}^{a}=\frac{B}{3}-L_{\mu}$  &  \  $K_{L_{\mu}}<1$ \\ \hline 
(3)  & \ \  $P_{1}^{a}=\frac{B}{3}-L_{\tau}$  &  \  $K_{L_{\tau}}<1$  \\ \hline 
(4)  & \ \  $P_{2}^{a}=L_{e}-L_{\mu}$  & \   $K_{L_{e}-L_{\mu}}<1$  \\ \hline	
(5)  & \ \ $ P_{2}^{b}=L_{e}-L_{\tau}$  & \   $K_{L_{e}-L_{\tau}}<1$  \\ \hline	
(6)  & \ \  $P_{2}^{c}=L_{\mu}-L_{\tau}$  & \   $K_{L_{\mu}-L_{\tau}}<1$   \\ \hline		
(7)  & \ \ $P_{3}^{a}=\frac{B}{3}+L_{e \mu}$ & \   $K_{L_{e \mu}}<1$  \\ \hline 
(8) & \ \ $P_{3}^{b}=\frac{B}{3}+L_{e \tau}$ & \   $K_{L_{e \tau}}<1$   \\ \hline 
(9) & \ \ $P_{3}^{c}=\frac{B}{3}+L_{\mu \tau}$ & \   $K_{L_{\mu \tau}}<1$   \\ \hline  
\end{tabular}
\caption{In the $SU(2)_{L}$ triplet Higgs model, all ten possible $U(1)$ charges 
and the conditions to enhance the approximate $U(1)$ charges are shown.
$L_{e \mu}$ is defined as $L_{e \mu}=L_{\tau}-L_{e}-L_{\mu}$.
$L_{\mu \tau}$ and $L_{e \tau}$ are defined similarly. 
The condition $K_{L_{e}}<1$ means the inequality, $\Gamma_{L_{e}}<H |_{T=M}$,
where we write all the processes which violate the charge $L_{e}$ as $\Gamma_{L_{e}}$.
The other conditions have similar meanings. \label{ap}}
\end{table}

\section{Summary}
We obtain the condition to protect the baryon number in the 
$SU(2)_{L}$ triplet Higgs model in (\ref{res}). The obtained condition  (\ref{res}) gives
the upper limits for either of the two coupling constants, $f^{\alpha \beta}$ and $A$,
which are newly introduced in addition to the standard model. 
When the difference of the lepton flavor, $e, \mu, \tau$, is considered,
the condition to protect the baryon number (\ref{res}) is modified into the condition (\ref{res2}).
In paper \cite{Has3}, we solve the Boltzmann equation and obtain
the condition which coincides with the condition (\ref{res}).
We further require in the paper that the model is compatible with the recent results of the neutrino
oscillation experiments and the 
Wilkinson Microwave Anisotropy Probe, and the constraints on the $\rho$ parameter imposed by 
the CERN LEP. We finally obtain the allowed region of the parameters in the model.

\vspace*{1cm}
\begin{center}
{\bfseries ACKNOWLEDGMENTS}
\end{center}  
This research was supported by the Science and Engineering Research Canada.

\end{document}